%% file: IEEE-conference-second-major.tex
\documentclass[journal]{IEEEtran}
\usepackage{cite}
\usepackage{amsmath,amssymb,amsfonts}
\usepackage{algorithmic}
\usepackage{graphicx}
\usepackage{textcomp}
\usepackage{xcolor}
\usepackage{comment}
\usepackage{url}
\usepackage{graphicx} 
\usepackage{multirow}
\usepackage{array}
\usepackage[caption=false,font=normalsize,labelfont=sf,textfont=sf]{subfig}
\usepackage{stfloats}
\usepackage{verbatim}

\def\BibTeX{{\rm B\kern-.05em{\sc i\kern-.025em b}\kern-.08em
    T\kern-.1667em\lower.7ex\hbox{E}\kern-.125emX}}
\usepackage{balance}
\usepackage{graphicx}
\usepackage{subcaption}

\begin{document}
\raggedbottom
\title{Graph Neural Networks for O-RAN Mobility Management: A Link Prediction Approach}

\DeclareRobustCommand{\IEEEauthorrefmark}
[1]{\smash{\textsuperscript{\footnotesize #1}}}

\author{
    \IEEEauthorblockN{
        Ana Gonzalez Bermudez\IEEEauthorrefmark{1},
        Miquel Farreras\IEEEauthorrefmark{1},
        Milan Groshev\IEEEauthorrefmark{1},
        José Antonio Trujillo \IEEEauthorrefmark{2},
        Isabel de la Bandera \IEEEauthorrefmark{2},
        Raquel Barco \IEEEauthorrefmark{2}\\
    }
    \IEEEauthorblockA{
        \IEEEauthorrefmark{1}The Laude Technology Company, S.L. 28037 Madrid, Spain \\
        \IEEEauthorrefmark{2}Telecommunication Research Institute (TELMA) University of Malaga, E.T.S.I. de Telecomunicacion, Bulevar Louis Pasteur 35, 29010, Malaga, Spain
    }
}

\maketitle 

\markboth{IEEE Vehicular Technology Magazine,~Vol.~X, No.~X, January~2025}%
{Shell \MakeLowercase{\textit{et al.}}: Bare Demo of IEEEtran.cls for IEEE Journals}

\begin{abstract}

Mobility performance has been a key focus in cellular networks up to 5G. To enhance Handover (HO) performance, 3GPP introduced Conditional Handover and Layer 1/Layer 2 Triggered Mobility (LTM) mechanisms in 5G. While these reactive HO strategies address the trade-off between HO failures and ping-pong effects, they often result in inefficient radio resource utilization due to additional HO preparations. To overcome these challenges, this article proposes a proactive HO framework for mobility management in O-RAN, leveraging user-cell link predictions to identify the optimal target cell for HO. We explore various categories of Graph Neural Networks (GNNs) for link prediction and analyze the complexity of applying them to the mobility management domain. Two GNN models are compared using \textcolor{black}{an anonymized real-world dataset and a synthetic dataset obtained with Keysight's EXata emulation tool. The results show the models' ability} to capture the dynamic and graph-structured nature of cellular networks. \textcolor{black}{Finally, we present key considerations for real-world deployment that outline future steps to enable the integration of GNN-based link prediction for mobility management in O-RAN networks.}
\end{abstract}

\begin{IEEEkeywords}
Graph Neural Networks, O-RAN, Handover, Mobility Management, Link Prediction.
\end{IEEEkeywords}

\section{Introduction}

Mobility management is a fundamental feature of cellular networks, ensuring seamless service continuity during user mobility by minimizing call drops, Radio Link Failures (RLFs) and unnecessary handovers (HOs), known as ping-ponging~\cite{3GPP_rel18}. This management can be carried out through different HO mechanisms, which are responsible for decision-making and message handling in the \textcolor{black}{Radio Access Network (RAN)}. Traditionally, mobility management follows a reactive HO mechanism, where HO decisions rely on predefined network parameters. However, determining optimal mobility parameters remains complex due to the trade-off between HO failures and ping-pong effects. Additionally, the dynamic communication environment often limits reactive techniques. These challenges are further exacerbated in 5G networks, where densification, higher frequencies, and stringent 
URLLC requirements demand more sophisticated mobility management solutions~\cite{ggn-ra}.

The advent of Open Radio Access Network (O-RAN) has revolutionized the RAN by leveraging Artificial Intelligence (AI) or Machine Learning (ML) to enhance network performance and flexibility. These AI/ML models, deployed at the edge and centralized cloud, enable intelligent decision-making and automation in managing modern cellular networks. Mobility management is a key O-RAN use case, and recent studies have explored proactive HO mechanisms based on user trajectory~\cite{yaj2022proactive} and User Equipment (UE) measurements~\cite{zaman2022come}. 
ML models such as sequence-to-sequence (Seq2Seq) and Long Short-Term Memory (LSTM) show promise in capturing long-term dependencies between UE and cell metrics.
However, these models process data sequentially, failing to leverage the graph-structured nature of communication networks, limiting generalization and reducing predictive performance on unseen networks and mobility patterns.

To address the challenges of modeling dynamic network topologies, Graph Neural Networks (GNNs) have become increasingly relevant in both academic and industrial contexts~\cite{STC, aws}. GNNs operate on graph-structured data by iteratively aggregating information from neighboring nodes, making them suitable for tasks such as node or edge regression and link prediction. Prior work has applied GNNs to regression problems in mobility management, such as predicting signal quality or load-related metrics~\cite{orhan2021}. Although these outputs can support HO optimizations, the models stop at intermediate predictions, with the estimation of HO targets remaining unexplored.

In this article, we study the capabilities of GNN link prediction for mobility management and how to overcome the technical challenges when adapting these ML models as proactive HO mechanism. This work is intended as an initial exploration, aimed at demonstrating the feasibility of using GNN-based link prediction to anticipate HOs in O-RAN systems. We propose next-cell prediction, a proactive HO mechanism that leverages a GNN-based link prediction model to forecast the upcoming cell connection for each user. We provide an in-depth analysis and model comparison of two categories of GNN link prediction models and analyze the complexity of applying them to the mobility management problem. Sec.~\ref{sec:backgorund} reviews the O-RAN framework, current HO mechanisms and the role of GNNs, while also summarizing recent advancements in proactive HO mechanisms. Sec.~\ref{sec:problem} outlines the problem formulation, proposes an O-RAN framework for next-cell prediction, and analyzes in-depth the challenges of applying GNNs for mobility management. Sec.~\ref{sec:results} compares the performance of prediction accuracy of subgraph-based and autoencoder-based GNN link prediction models, trained \textcolor{black}{on real-world and synthetic datasets}. \textcolor{black}{Sec.~\ref{sec:discussion} summarizes the main considerations for the real-world deployment of GNN-based link prediction for mobility management. Finally, Sec.~\ref{sec:conclusion} concludes the article with key findings and future directions.}

\section{Background}
\label{sec:backgorund}

\subsection{O-RAN}
\textcolor{black}{O-RAN is an industry effort to open and disaggregate the RAN, enabling multi-vendor interoperability, software driven control, and AI/ML assisted automation. Its architecture centers on two RAN Intelligent Controllers (RICs), the Non-Real-Time (Non-RT) RIC hosted in the Service Management and Orchestration (SMO) system, and the Near-Real-Time (Near-RT) RIC  close to the RAN. rApps run in the Non-RT to train models and issue policies over the A1 interface, while xApps run in the Near-RT RIC to apply those policies together with measurements received over the E2 interface. Network configuration and performance data flow between the SMO and RAN through the O1 interface. In parallel, functional disaggregation separates the Central Unit (CU) and the Distributed Unit (DU) and further splits the CU into control plane (CU-CP) and user plane (CU-UP) parts, which simplifies scaling, upgrades, and vendor mixing within a gNB deployment.}

\subsection{Handover mechanisms in O-RAN}
\begin{figure}[h]
    \centering
    \includegraphics[width=0.65\columnwidth]{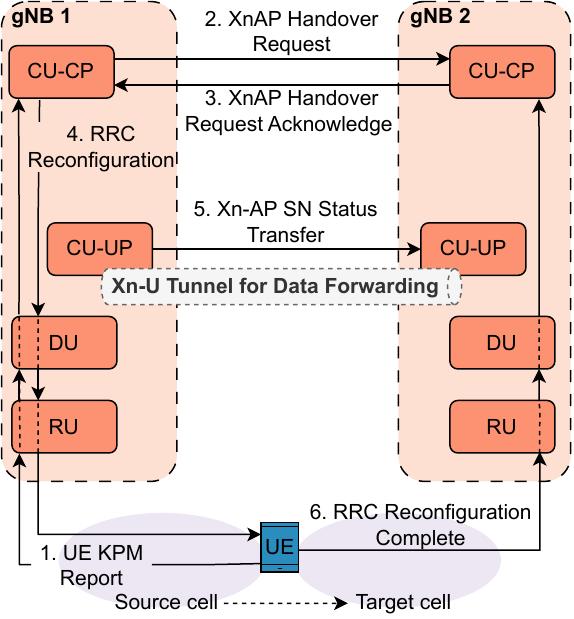}
    \vspace*{-1ex}
    \caption{Handover process in cellular networks.}
    \label{fig:handover}
    \vspace*{-1ex}
\end{figure}

This work addresses inter-gNB HO in O-RAN, utilizing the Xn interface for transitions between different gNBs, as shown in Fig.~\ref{fig:handover}. The baseline HO process involves the UE measuring neighboring cell metrics, such as Signal-to-Noise Ratio (SNR), and reporting these to its source gNB (1). If the source signal degrades and neighbor's improves, the source gNB initiates the HO procedure. This involves a sequence of steps: an XnAP Handover Request is sent to the target gNB (2), the target gNB replies with an acknowledgment (3), the source gNB then instructs the UE to reconfigure its connection (4), a temporary user-plane tunnel is set up between source and target CU-UPs for data transfer, and finally the UE confirms the reconfiguration to the target gNB (6).

O-RAN facilitates the integration of advanced AI/ML-based solutions in the RICs via rApps and xApps. The distinction lies in their response time restrictions: rApps operate with response times greater than 1 second, while xApps require response times between 10 ms and 1 second. In the standard ~\cite{oran_wg3_uc}, O-RAN introduces the use of these AI/ML solutions to implement proactive HO mechanisms for mobility management. The Non-RT RIC trains the AI/ML models using collected metrics via the O1 interface and then the deployed rApp sends A1 policies to the Rear-RT RIC. The xApp deployed in the near-RT RIC subsequently uses these A1 policies, alongside E2 measurements, to apply real-time configuration updates, facilitating proactive and adaptive HO decisions.

\subsection{GNNs for link prediction in mobility management}

GNNs have emerged as powerful tools for learning over structured data. Their ability to process graph-structured data, by aggregating information from neighboring nodes, allows them to capture both local and global graph properties. Among the many tasks GNNs address (e.g., regression, classification), link prediction stands out as particularly relevant for scenarios where the goal is to infer potential or future connections between entities.

Several GNN architectures have been developed specifically for link prediction. Among them, encoder-decoder models such as Graph Autoencoders (GAE), which produce fixed embeddings. While Variational Graph Autoencoders (VGAE) model distributions, capturing uncertainty seen in dynamic mobility settings, subgraph-based methods focus on local neighborhoods around candidate links to capture short-range patterns, being particularly effective in sparse or evolving graphs. SEAL, for instance, applies GNNs to these subgraphs in a supervised manner, enabling fine-grained link prediction. Matrix factorization models like ComplEx embed nodes and relations into complex vector spaces, enabling them to capture asymmetry. Table~\ref{table:models} summarizes the most important GNN models for link prediction, outlining their inputs, outputs, benefits, and limitations.

\input{tables/model-summary}

\subsection{Existing solutions for mobility management}
Recent AI/ML approaches for mobility management~\cite{drl-2} \textcolor{black}{include} Seq2Seq models for intra-frequency HO \textcolor{black}{optimizations using} UE trajectories~\cite{yaj2022proactive}, LSTM models for next-cell prediction with mobility context~\cite{zaman2022come} and \textcolor{black}{Deep Reinforcement Learning (DRL) methods that learn policies from interactions~\cite{drl-1, drl-2}}. \textcolor{black}{These families of models capture temporal patterns and can adapt to changing traffic and load. In this article, we study a graph-based formulation that operates on the network structure, with the goal of assessing whether link prediction can support proactive decisions in an O-RAN. We propose this as complementary to DRL and recurrent models, as each family targets different aspects of the problem.}

Cellular networks naturally form graphs of interdependent nodes with features, making GNNs suitable for solving network problems. They have been applied to radio resource management~\cite{ggn-ra}, 5G signal inference~\cite{5GNN} and traffic prediction~\cite{STC}. In the case of link prediction, models must account for factors like interference and load, which introduce temporal variability. Recent studies combine GNNs with RL, as in~\cite{orhan2021}, where connection management is performed as a graph optimization task. 
Industry efforts such as Amazon's GNN-based mobility solution~\cite{aws} highlight the practical interest in this domain, though technical details remain unpublished. Furthermore,~\cite{hasan2024} combines GNNs and time-series transformers to predict RLFs, but focuses on link reliability than mobility management.

Despite these advancements, proactive HO management using GNNs remains underexplored. The complexity of dynamic graph-structured networks presents an opportunity for GNN-based methods adapted to mobility and operational constraints.

\section{Next-cell prediction for mobility management}
\label{sec:problem}

\subsection{Problem formulation}
The baseline HO mechanism can lead to signaling overhead and delays, due to necessary re-transmissions which cause the HO to be triggered too late. Additionally, the interruption time is typically between 50-90 ms, often too long for URLLC services. Network-triggered HO mechanisms like conditional HO and L1/L2 Triggered Mobility (LTM), although designed to improve this interruption time, can lead to ping-ponging. Moreover, triggers based on low-level metrics may ignore the overall network quality or congestion levels, potentially leading to suboptimal HO. Traditional ML proactive approaches, although promising, may struggle with adaptability and generalization in dynamic environments~\cite{barzizza2024recent}. 

To address these limitations, we propose a GNN-based link prediction method for proactive mobility management in cellular networks. This approach distinguishes itself from traditional reactive or sequential methods by enabling direct next-cell prediction for UEs, thereby aiming to reduce signaling overhead and HO delays. By leveraging the spatial and graph-structured context of cellular networks, our models estimate future user-cell links. This initial study demonstrates the feasibility of supporting earlier HO decisions, with the potential to minimize ping-pong effects and improve HO timing. The autoencoder-based model, in particular, suggests a scalable solution for enhancing HO performance and user experience within O-RAN architectures.

\subsection{Next-Cell prediction rApp - A Walkthrough}

\begin{figure}
    \centering
    \includegraphics[width=1\columnwidth]{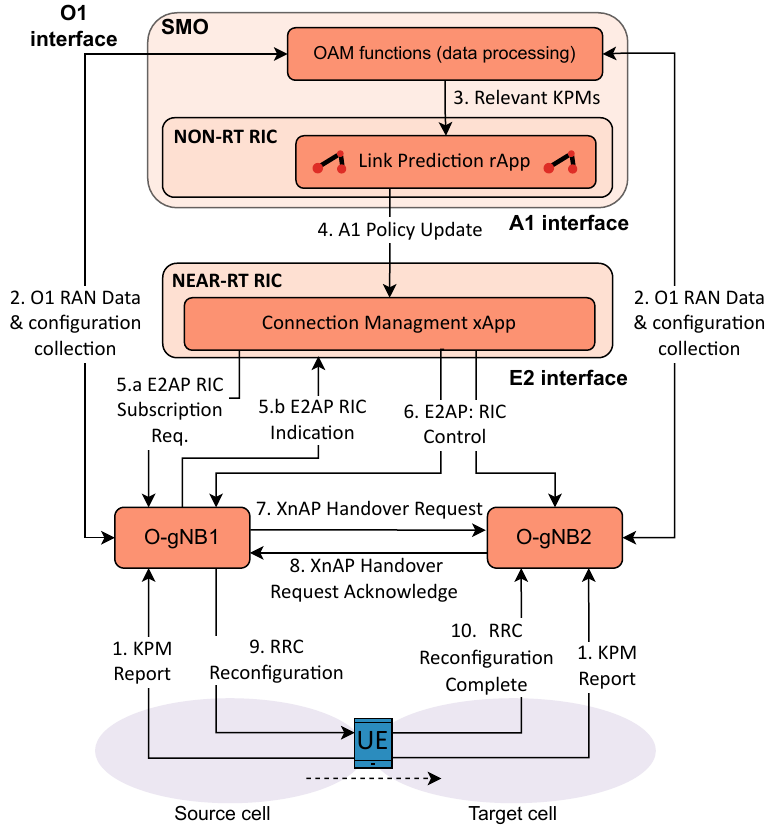}
    \vspace*{-1ex}
    \caption{Next-cell prediction framework for proactive HO.}
    \label{fig:architecture}
    \vspace*{-3ex}
\end{figure}

The proposed O-RAN mobility management solution is shown on Fig.~\ref{fig:architecture} and operates by integrating the next-cell prediction rApp into the Non-RT RIC. The workflow starts with UE collecting and transmitting measurement reports (1). Subsequently, both the serving and neighboring gNBs transmit RAN data and configurations, including L1/L2/L3 metrics, to the SMO Operations, Administration, and Maintenance (OAM) functions via the O1 interface (2). These OAM functions have a baseline measurement configuration to select the data of interest, and allows its access to the Non-RT RIC (3). A pre-trained \textcolor{black}{(with prior knowledge of the network)} GNN-based link prediction model, hosted as an rApp evaluates the collected RAN data to forecast the upcoming cell connections. Based on these predictions, the rApp generates A1 Policy update messages and transmits them to the Connection Management xApp via the A1 interface (4). This xApp, based on the the A1 policies and the E2 measurement metrics obtained from the E2 nodes (5.a and 5.b), generates RIC Control messages (6) to initiate radio resource re-allocation and HO request (7 and 8), mirroring the standard HO process from Fig.~\ref{fig:handover}. Crucially, the only HO latency introduced by this solution is attributed to O1 data collection, model inference (evaluated in Sec. IV.D), and policy generation, all of which are low-overhead operations. This design choice is fundamental to maintaining efficiency and adhering to the stringent timing requirements inherent in O-RAN deployments.

\subsection{Challenges of applying GNN models for mobility management}
\label{sec:gnn_challenges}

Applying GNN models for next-cell prediction in mobility management presents distinct challenges, primarily due to the dynamic, heterogeneous, and imbalanced characteristics of real-world cellular networks.

\subsubsection{\textbf{Data heterogeneity}}
Cellular networks are heterogeneous, comprising diverse node types (e.g., gNBs, cells, UEs) and different edge types (e.g., radio links, HO transitions, control interfaces). This heterogeneity is critical for accurately modeling mobility patterns, as HO decisions depend on both user-cell associations and inter-cell relationships. However, standard GNN models for link prediction are designed for homogeneous graphs, which prevents the effective representation of these complex network structures.

\subsubsection{\textbf{Data imbalance}}
Mobility management scenarios exhibit pronounced node imbalance, as the number of UE nodes significantly exceeds the number of cell nodes. This can lead to low connectivity and sparse graphs, which can significantly impact GNN models. The imbalance affects message-passing GNN models, since low-degree nodes may receive insufficient aggregated information, limiting predictive accuracy. For example, in next-cell prediction, imbalanced representations can bias the model toward frequent HO targets, reducing sensitivity to less common transitions.
The imbalance also complicates prediction thresholds, as the models produce probabilistic scores rather than binary outputs. A default threshold of 0.5 is often unsuitable in imbalanced scenarios~\cite{zou2016finding}.

\subsubsection{\textbf{Edge features}}
Edge features play a central role in modeling HO behavior, as they encode contextual information (e.g., signal strength, timing, HO frequency) critical to predicting next-cell transitions. However, many baseline GNN models aggregate node embeddings without considering edge attributes, limiting their effectiveness in O-RAN scenarios.

\section{Model evaluation}
\label{sec:results}
This section evaluates two GNN models: VGAE based and SEAL. VGAE captures global embeddings, while SEAL focuses on localized subgraphs, providing a complementary comparison for mobility management. The following subsections detail the , model creation, training methodology, and experimental setup.

\subsection{\textcolor{black}{Used datasets}}

\textcolor{black}{To validate the model performance we used two datasets: \emph{i)} an open-source Real-World dataset (RW dataset) from Huawei Technologies\footnote{https://snap.stanford.edu/data/telecom-graph.html}, and 
\emph{ii)} a synthetic EMulated dataset (EM dataset) generated using Keysight 5G network emulator  EXata\footnote{https://www.keysight.com/fr/en/product/SN100VGLA/exata-network-modeling-5g.html}. RW was selected for its scale, structure, and operational origin, providing a strong foundation for an initial GNN-based link prediction model development. The small-scale EXata dataset (EM), was generated for validation of the models generalization capabilities across diverse network scenario with real radio metrics. The datasets main characteristics are:
\begin{itemize}
    \item RW is composed of 35 million edges and 169,000 nodes (100,000 UEs and 69,000 cells), which allows the analysis of the models' scalability. All nodes and edge have features associated: UEs have 1,000, cells have 300, and edges have 37. However, for privacy reasons, all node and edge features are anonymized, represented as random properties normalized between 0 and 1, which limits the data analysis. 
    \item EM is composed of 449,152 edges and 101 nodes (70 UEs and 31 cells). The emulator allows the collection of L1/L2 UE radio metrics. Performing feature engineering on the collected data and excluding the highly correlated features, as UE-cell edges we used: RSRP, RSRQ, Transport Blocks, Available Resources Blocks, Packet Size, MCS, Signal Power Uplink, and Signal Power Downlink. 
\end{itemize}
}

\textcolor{black}{From the raw data, we create the same graph data structure for both datasets. Due to computational complexity of processing RW and the scale differences, initial development is conducted on a subset of 985 cells and 10,000 users, with further for gradual scaling. Also, RW limits its interpretability due to a lack of information on multiple edges between the same pair of UE-cell. To manage potential redundancy, only one edge per user-cell pair (the one with the lowest ID) was used, reducing the total edges to approximately 7 million. Same procedure is then used in EM, reducing the edges to 489. This way, RW has each UE connected to 69 cells and each cell to 1,000 UEs, creating 1,000 isolated sub-networks without inter-connectivity. EM has each UE connected with different number of cells, ranging from two to six. These structures yields sparse network densities of 0.00049 and 0.09683, respectively.}

\subsection{Model creation}
The link prediction task is formally defined over an attributed graph
\(
G = (V, E, X),
\)
where \(V\) denotes the set of nodes, \(E\) the set of edges, and \(X\) represents node and edge features. The objective is to predict the probability of a link existing between pairs of nodes \((u, v)\), corresponding to the likelihood that a user will HO to a given target cell.

Model adaptations were necessary due to the graph's specific nature and the solution's objectives. As the dataset comprises two node types connected through a single hop, the model design employs a single layer for data transmission between neighbors. Incorporating additional layers would result in redundant information transmission and lead to over-smoothing and learning inefficiencies.

To address each challenge of GNN models for mobility management, the following adaptations are implemented:

1) \textbf{Data heterogeneity}: The heterogeneous topology is transformed into an equivalent homogeneous graph. It is carried out unifying UEs and cells in a single type of node, assigning a unique ID for each one while preserving relevant information, such as features and connections.
    
2) \textbf{Data imbalance}: Subgraph-based GNNs mitigate global data imbalance by capturing specific subgraph structural patterns. However, autoencoder-based models are particularly affected due to their reliance on neighboring embeddings, which can decrease overall link prediction accuracy. An attention-based mechanism, Graph Attention Network (GAT), is used to mitigate this by prioritizing important neighbors during aggregation, thereby improving representation for underrepresented nodes. Additionally, due to data imbalance, prediction thresholds in both models require empirical tuning to identify decision boundaries that balance true positives and true negatives, ensuring stable performance metrics (e.g., recall and accuracy).

3) \textbf{Edge features}: For subgraph-based models, edge features are incorporated at the initial stage of subgraph processing, being encoded with node features. GAT layers enable the inclusion of edge features in the autoencoder-based models.

Furthermore, specific adaptations are made for VGAE and SEAL models.

\subsubsection{VGAE}

The VGAE encoder's message-passing operation, utilizing a single GAT layer (expressed as Eq.~\ref{eq:vgae}), where \(\mathbf{H}^{(l-1)}\) is the previous layer node features, \(\hat{\mathbf{A}}\) is the symmetrically normalized adjacency matrix, \(\mathbf{H}^{(l)}\) is the output of the current layer, \(\mathbf{W}^{(l)}\) is a trainable weight matrix, and \(\sigma\) is a ReLu activation function. The GAT layer's output is processed by two GCN layers to calculate the mean (\textit{$\mu$}) and log standard deviation (\textit{logstd}), capturing uncertainty in the graph structure. An inner product decoder then calculates edge probabilities, with a threshold applied for the classification of edges. This combination of VGAE and GAT focuses on influential nodes while adapting to dynamic changes, making it ideal for next-cell prediction. 

\begin{equation}
\mathbf{H}^{(l)} = \sigma\left( \hat{\mathbf{A}} \mathbf{H}^{(l-1)} \mathbf{W}^{(l)} \right),
\label{eq:vgae}
\end{equation}

\subsubsection{SEAL}

In SEAL, the labeled subgraphs are processed by a GCN-based encoder, which learns a compact subgraph embedding. This enables the model to capture both structural and relational information from the dataset. The message-passing operation in SEAL can be expressed as Eq.~\ref{eq:seal}, where \(\mathbf{H}_i^{(l)}\) is the hidden representation of node \(i\) in layer \(l\), \(\mathbf{W}^{(l)}\) is the trainable weight matrix at layer \(l\), and \(\sigma\) is a non-linear ReLu activation function.

\begin{equation}
\mathbf{H}_i^{(l)} = \sigma\left( \sum_{j \in \mathcal{N}(i)} \mathbf{W}^{(l)} \mathbf{H}_j^{(l-1)} \right),
\label{eq:seal}
\end{equation}

\subsection{Model training}
A systematic training process was applied to optimize model performance. The datasets was divided into 80/10/10 training, validation, and test sets using a custom split function to ensure consistent, structure-preserving partitions. This split generates positive (existing) and negative (non-existing) edges for each set, which are necessary for contrastive supervision during training. This custom approach prevents potential issues associated with automatic split generation functions (e.g., PyTorch \textit{RandomLinkSplit}), such as including existing edges as negative, and duplicated negative edges between different sets, which negatively impact model performance~\cite{Zhou2023}.

VGAE and SEAL use labeled data for supervised training with Kullback-Leibler (KL) divergence and cross-entropy loss, respectively. The Adam optimizer was used for its efficiency in handling large and sparse datasets. Both models underwent fine-tuning on parameters such as weights, learning rate, latent representation size. During training, model performance was evaluated using Area Under the ROC Curve (AUC), Average Precision (AP), Precision, Recall, F1-score, Accuracy, and Matthews Correlation Coefficient (MCC). To address potential overfitting, early stopping was applied to both models, and L2 regularization in SEAL.

\subsection{Performance comparison}
\begin{figure*}
    \centering
    \includegraphics[width=\textwidth]{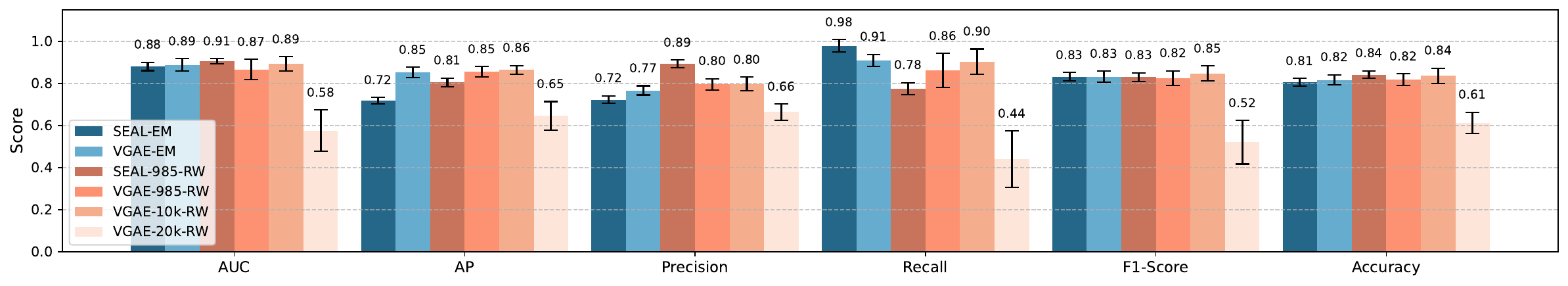}
    \vspace*{-1ex}
    \caption{\textcolor{black}{Performance metrics obtained with VGAE and SEAL, along with different dataset types and sizes.}}
    \label{fig:metrics_all}
\end{figure*}

Fig.~\ref{fig:metrics_all} shows the mean metrics values obtained from 10 executions of each \textcolor{black}{model-dataset combination. The blue bars correspond to the results with the EM dataset with SEAL (SEAL-EM) and VGAE (VGAE-EM). The orange bars represent SEAL and VGAE applied to the smallest partition of RW, with 985 cells and 10,000 UEs (SEAL-985-RW and VGAE-985-RW), and the VGAE results when scaling the dataset to 10,000 cells and 100,000 UEs (VGAE-10k-RW) and to 20,000 cells and 1,000,000 UEs (VGAE-20k-RW).}

\textcolor{black}{Using the EM dataset, VGAE-EM outperforms SEAL-EM in AUC (0.81\%), AP(18.63\%), Precision (6.08\%), F1-score (0.1\%) and Accuracy (1.36\%), whereas SEAL-EM outperforms VGAE-EM in Recall (7.53\%). Meanwhile, using the 985-RW dataset, VGAE-985-RW achieves higher AP (6.21\%) and Recall (11.23\%) than SEAL-985-RW, but SEAL-985-RW outperforms VGAE-985-RW in AUC (4.5\%), Precision (12.3\%), F1-score (10.09\%) and Accuracy (0.61\%). The differing outcomes observed in the models' performance, depending on the used dataset, arise from their inherent differences in density and size.
SEAL shows lower performance with the EM, as the low dataset size produces small subgraphs that limit its learning capacity. In contrast, VGAE enhances its performance with EM, as it is almost 200 times denser than RW, which particular benefits to autoencoder-based models capturing edges within a broader context. This is the reason for VGAE worsening with RW, generating more False Positives (FP) due to the sparsity of the graph. However, with the larger dataset SEAL achieves stronger performance in FP-sensitive metrics.} 

\textcolor{black}{Fig.~\ref{fig:metrics_all} also illustrates VGAE scalability with increasing graph size.} 
SEAL could not run on these larger datasets due to the significant computational cost associated with subgraph creation. VGAE performance improved overall when \textcolor{black}{scaling from VGAE-985-RW to VGAE-10k-RW, with} Recall showing the highest growth (4.91\%), followed by AUC (3\%), Accuracy (2.2\%) and F1-score (2.18\%). This reflects an increase in correctly predicted edges, due to improved generalization from training the model on a broader variety of subgraph structures and interactions. As a result, the model learns to assign more accurate attention to nodes during training. However, \textcolor{black}{VGAE-20k-RW} model performance deteriorates. This occurred because the dataset maintained the same number of UEs, but duplicated cells, leading to a drastic increase in node density, which likely affects the attention mechanism and increases FPs (key limitation of the model).

\begin{figure}
    \centering
    \includegraphics[width=0.8\columnwidth]{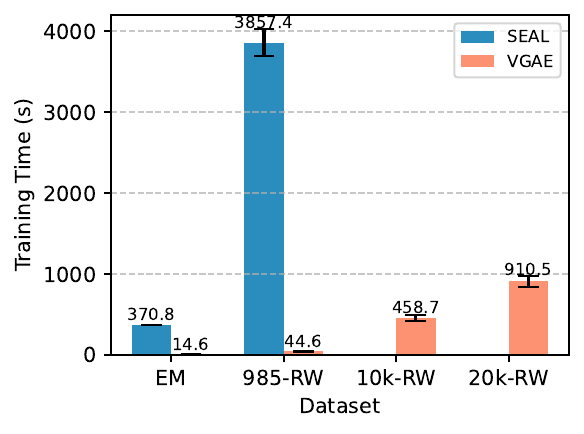}
    \vspace*{-1ex}
    \caption{\textcolor{black}{Training time of VGAE and SEAL, along with different dataset types and sizes.}}
    \label{fig:train_time}
\end{figure}

Fig.~\ref{fig:train_time} shows the models training times. VGAE  outperforms SEAL in execution speed, achieving a 98.84\% reduction in training time \textcolor{black}{for 985-RW and a 38.5\% for EM} due to its direct graph processing. This computational burden prevented SEAL from executing on the two large datasets. VGAE also demonstrates faster convergence during training, achieving satisfactory results within less than 50 epochs \textcolor{black}{in both datasets}. This rapid convergence, coupled with its simpler architecture, enables efficient training, even on larger datasets. Furthermore, VGAE's training time increases linearly with the graph size as it performs embeddings for each node of the network.

\begin{figure}
    \centering
    \includegraphics[width=0.8\columnwidth]{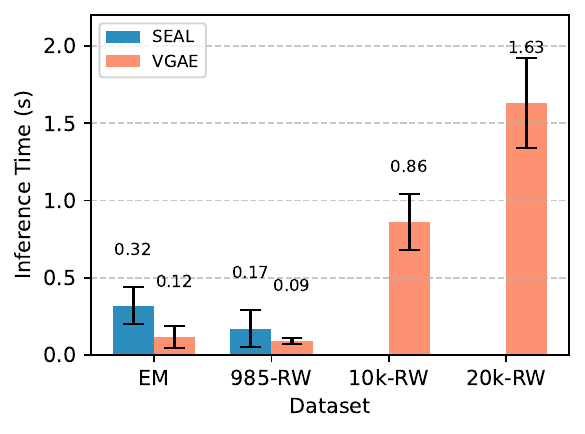}
    \vspace*{-1ex}
    \caption{\textcolor{black}{Inference time of VGAE and SEAL, along with different dataset types and sizes.}}
    \label{fig:inf_time}
\end{figure}

Fig.\ref{fig:inf_time} illustrates VGAEs reduction in SEAL's inference time \textcolor{black}{for 985-RW and for EM}, also attributed to its direct execution without subgraphs. VGAEs inference time, similar to its training time, also grows linearly with dataset size.  Crucially, all model inference times are below or close to 1 second, demonstrating their high efficiency and suitability for O-RAN rApp deployments, which are characterized by operational response times greater than 1 second.

The findings demonstrate the efficacy of graph-based methodologies for link prediction in \textcolor{black}{different} networks, even without feature information \textcolor{black}{(RW dataset) or temporal processing}, relying solely on graph structure for learning. While prior studies used GNN-based regressions models for HO estimation (e.g., ~\cite{hasan2024} for radio link failure and~\cite{yaj2022proactive} for user mobility), no prior work directly applied GNN models for link prediction in this context.  Although not directly comparable, our proposal achieved an F1-score 5\% higher (84\%) than~\cite{hasan2024} (79\%), using only graph structure information. Similarly, \cite{yaj2022proactive} reported 85\% accuracy with nodes features processing, improvable to 90\% by including user's historical mobility data. This suggests that including temporality and features would enhance our model performance, which already demonstrates high efficiency using only the graph structure for training. 

\section{\textcolor{black}{Considerations for real-world deployment}}\label{sec:discussion}
\textcolor{black}{Having demonstrated the feasibility of using GNNs for HO prediction, deploying this solution in a live O-RAN environment requires addressing several key challenges. The following considerations are crucial for evolving the proposed framework from a proof-of-concept into a robust, scalable, and adaptable real-world tool.}

\subsection{\textcolor{black}{Enhancing Model Architecture for Dynamic Networks}}
\textcolor{black}{
A primary consideration for real-world deployment is the evolution of the model to handle the dynamic and temporal nature of cellular networks. The VGAE and SEAL models are transductive. They are trained on a fixed graph and struggle to generalize to new UEs entering the network or to different network environments without retraining. Furthermore, a key challenge is modeling the continuous evolution of a UE's connection as it moves, which constantly changes the state of its wireless channel to the serving cell. This can be represented in our framework by creating multiple, time-stamped edges between the same UE-cell pair.
\textcolor{black}{Thereby, the model can be fed with the data related with the edge evolution over the time, and predict the edge to a new cell.} Future work should explore the adaptation of inductive models, such as Spatio-Temporal GNNs, which learn transferable patterns applicable to unseen nodes. This type of models can natively learn from the sequence of the channel changes to better understand a UE's trajectory and make more context-aware predictions.
}
\subsection{\textcolor{black}{Addressing Data-Centric Challenges}}
\textcolor{black}{
The unique characteristics of cellular network data introduce further challenges that must be managed for real-world deployment. Cellular graphs are naturally sparse, with  UEs connected to the serving cell. This low graph connectivity can hinder the message-passing mechanism essential for GNNs, as nodes have limited information to aggregate from their neighbors. A potential solution is to enrich the graph structure by incorporating higher-level network nodes, such as CUs and DUs, to create a more densely connected graph that facilitates richer information flow. Another critical data-centric task is the generation of negative edges for training, as a naive random sampling approach can introduce flaws like including existing links as negative examples or duplicating samples, leading to overfitting and biased evaluation. Future implementations should use a structure-aware sampling strategy to generate more realistic negative samples. \textcolor{black}{Finally, GNNs demonstrate that they can operate effectively on anonymized data by learning from the graph structure. However, GNNs present a "black box" when it comes down to the model explainability. This limitation can be addressed by integrating specialized explainability GNN frameworks, such as GNNExplainer or GraphSHAP designed to interpret GNN predictions.}}

\subsection{\textcolor{black}{Ensuring Scalability and Operational Viability}}
\textcolor{black}{For integration of our solution into a real-world scenario, scalability is the central requirement. The solution must remain computationally efficient, operate within the available budget, and grow with network size. GNN architectures differ in cost and accuracy. Subgraph based models such as SEAL can be precise on small datasets, yet their per sample search and training cost rises quickly, which limits scale. Models such as VGAE are lighter and fit better for large graphs, although performance can drop in very dense areas. To meet operator needs, a modular approach can be used. The operator network can be partitioned into regions, for example dense urban zones with frequent HOs, suburban zones with moderate mobility, and high speed corridors with fast mobility. Each region runs a model and set of hyperparameters tailored to its conditions, which keeps inference latency and memory within target limits.}

\section{Conclusions} \label{sec:conclusion}
In this article we proposed a GNN-based link prediction solution for proactive mobility management in O-RAN. Using real-world and synthetic cellular data, we compared two GNN models: an autoencoder-based model and a subgraph-based model, with results showing that the autoencoder model offers superior speed and training efficiency, especially for larger datasets, achieving competitive predictive performance with high recall and average precision. While the subgraph model excelled in precision, the autoencoder model demonstrated better generalization and recall with larger datasets, though its performance decreased with high node density. Given the ability of the autoencoder-based model to deliver accurate next-cell predictions efficiently and its potential to address challenges in mobility management, we advocate for further exploration of this approach. Future research could refine and extend these findings, paving the way for designing innovative GNN-based HO mechanisms in future networks, marking a significant paradigm shift in mobility management.

\section*{Acknowledgment}

This project is partially funded by the European Union-NextGenerationEU within the Framework of the Project "Massive AI for the Open RadIo b5G/6G Network (MAORI)".

\bibliographystyle{IEEEtran}
\bibliography{IEEEabrv,IEEEsettings,ref}{}

\section*{Biographies}
\small{
  \textbf{Ana González Bermúdez} (M.Sc.'2022): Researcher at Laude.
  
  \textbf{Miquel Farreras} (M.Sc.'2019, Ph.D.'2024): Researcher at Laude.
  
  \textbf{Milan Groshev} (M.Sc.'2016, Ph.D.'2022): Senior Researcher at Laude.
  
  \textbf{José Antonio Trujillo} (M.Sc.'2020): Ph.D Candidate at UMA.

  \textbf{Isabel de la Bandera} (M.Sc.'2009, Ph.D.'2017): Associate professor at UMA.

  \textbf{Raquel Barco} (M.Sc.'1997, Ph.D.'2007): Full professor at UMA.
}

\end{document}

%% file: tables/model-summary.tex
\begin{table*}[ht!]
\centering
\renewcommand{\arraystretch}{1.5} 
\caption{Summary of link prediction models}

\begin{tabular}{|p{0.2cm}|p{1cm}|p{2.35cm}|p{2.35cm}|p{4.75cm}|p{4.75cm}|}
\cline{1-6}

\textbf{} & \textbf{Name} & \textbf{Input} & \textbf{Output} & \textbf{Strengths} & \textbf{Weaknesses} \\ \hline

\multirow{2}{*}{\raisebox{-1\height}{\rotatebox{90}{\textbf{Autoencoders}}}}

& \textbf{GAE} & 
Node features, edge features and edge indexes. & 
Predicted existence or non-existence of edges. & 
Its architecture is highly adaptable. Also, it is simpler and more efficient than other models.  &  
Unsupervised and deterministic model with a binary output, so no further analysis is allowed. \\ \cline{2-6}

& \textbf{VGAE} & 
Node features, edge features and edge indexes. & 
Predicted probabilities for each edge. & 
Captures uncertainty and variability in the data. Improves the adaptability in networks with changing user behavior. & 
Unsupervised model. Requires an additional process to determine an optimal threshold for the probabilities obtained. \\ \cline{2-6}

\hline
\multirow{2}{*}{\raisebox{-0.95\height}{\rotatebox{90}{\textbf{Subgraph-based}}}}

& \textbf{SEAL} & 
Enclosing subgraphs and node features. & 
Binary classification (link or no link between two nodes). & 
Focuses on local subgraphs, improving short-range predictions; supervised learning enhances performance. & 
Subgraph extraction increases computational cost; requires labeled data. \\ \cline{2-6}

& \textbf{ComplEx} & 
Node pairs and relation types (for multi-relational graphs). & 
Predicted probabilities of links between nodes. & 
Handles multi-relational graphs; captures asymmetric relationships effectively. & 
Limited structural information beyond embeddings; computationally intensive for large-scale graphs. \\ \hline

\end{tabular}
\label{table:models}
\vspace*{-3ex}%
\end{table*}